# Automated Chest CT Image Segmentation of COVID-19 Lung Infection based on 3D U-Net

Dominik Müller[1], Iñaki Soto Rey[1] and Frank Kramer[1]

[1] IT-Infrastructure for Translational Medical Research, Faculty of Applied Computer Science, Faculty of Medicine, University of Augsburg, Germany

**ABSTRACT**

**The coronavirus disease 2019 (COVID-19) affects billions of lives around the world and has a significant impact on public healthcare. Due to rising skepticism towards the sensitivity of RT-PCR as screening method, medical imaging like computed tomography offers great potential as alternative. For this reason, automated image segmentation is highly desired as clinical decision support for quantitative assessment and disease monitoring. However, publicly available COVID-19 imaging data is limited which leads to overfitting of traditional approaches. To address this problem, we propose an innovative automated segmentation pipeline for COVID-19 infected regions, which is able to handle small datasets by utilization as variant databases. Our method focuses on on-the-fly generation of unique and random image patches for training by performing several preprocessing methods and exploiting extensive data augmentation. For further reduction of the overfitting risk, we implemented a standard 3D U-Net architecture instead of new or computational complex neural network architectures. Through a 5-fold cross-validation on 20 CT scans of COVID-19 patients, we were able to develop a highly accurate as well as robust segmentation model for lungs and COVID-19 infected regions without overfitting on the limited data. Our method achieved Dice similarity coefficients of 0.956 for lungs and 0.761 for infection. We demonstrated that the proposed method outperforms related approaches, advances the state-of-the-art for COVID-19 segmentation and improves medical image analysis with limited data. The code and model are available under the following link: https://github.com/frankkramer-lab/covid19.MIScnn**

## 1. INTRODUCTION

The ongoing coronavirus pandemic has currently (4[th] of May 2020) spread to 187 countries in the world [1]. The World Health Organization (WHO) declared the outbreak as a "Public Health Emergency of International Concern" on the 30[th] of January 2020 and as a pandemic on the 11[th] of March 2020 [2,3]. Because of the rapid spread of severe respiratory syndrome coronavirus 2 (SARS-CoV-2), billions of lives around the world were changed. A SARS-CoV-2 infection can lead to a severe pneumonia with potentially fatal outcome [3–5]. Until now, there are 3,531,618 confirmed cases in total resulting in 248,097 deaths [1]. So far, there is neither an effective treatment for the infection, nor is there an effective prevention against it, such as a vaccination [3,4,6,7]. Additionally, the rapid increase of confirmed cases and the resulting estimated basic reproduction numbers show that SARS-CoV-2 is highly contagious [4,6,8]. Therefore, fast detection and isolation of infected persons are crucial in order to limit the spread of the virus. The WHO named this new disease "coronavirus disease 2019", short form: COVID-19.

The reverse transcription polymerase chain reaction (RT-PCR) was established as the standard approach for COVID-19 screening [2,4,6]. RT-PCR is able to detect the viral RNA in specimens obtained by nasopharyngeal swab, oropharyngeal swab, bronchoalveolar lavage or tracheal aspirate [2,4,6,7]. However, a variety of recent studies indicate that RT-PCR testing suffers from a low sensitivity, approximately around 71%, whereby repeated testing is needed for accurate diagnosis [9,10]. Furthermore, RT-PCR screening is time-consuming and has increasing availability limitations due to shortage of required material [10].

An alternative solution to RT-PCR for COVID-19 screening is medical imaging like X-ray or computed tomography (CT). The medical imaging technology has made significant progress in recent years and is now a commonly used method for diagnosis, as well for quantification assessment of numerous diseases [11–13]. Particularly, chest CT screening has emerged as a routine diagnostic tool for pneumonia. Therefore, chest CT imaging has also been strongly recommended for COVID-19 diagnosis and follow-up [9]. In addition, CT imaging is playing an important role in COVID-19 quantification assessment, as well as disease monitoring. COVID-19 infected areas are distinguishable on CT images by ground-glass opacity (GGO) in the early infection stage and by pulmonary consolidation in the late infection stage [6,9,14]. In comparison to RT-PCR, several studies showed that CT is more sensitive and effective for COVID-19 screening, and that chest CT imaging is more sensitive for COVID-19 testing even without the occurrence of clinical symptoms [9,10,12,14]. Notably, a large clinical study with 1,014 patients in Wuhan (China) determined that chest CT analysis can achieve 0.97 sensitivity, 0.25 specificity and 0.68 accuracy for COVID-19 detection [9].

Still, evaluation of medical images is a manual, tedious and time-consuming process performed by radiologists. Even though increasing CT scan resolution and number of slices resulted in higher sensitivity and accuracy, these improvements also increased the workload. Additionally, annotations of medical images are often highly influenced by clinical experience [15,16].

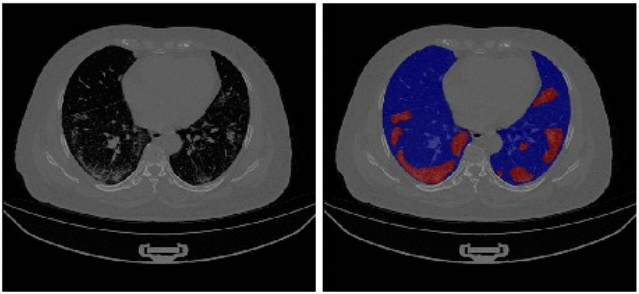

Figure 1: Visualization of COVID-19 infected regions in a chest CT. The left image is the unsegmented CT scan, whereas the right image shows segmentation of lungs (blue) and infection (red). The infected regions are distinguishable by GGOs and pulmonary consolidation in the lung regions. The image was obtained from the analyzed CT dataset [36].

A solution for these challenges could be clinical decision support systems based on automated medical image analysis. In recent years, artificial intelligence has seen a rapid growth with deep learning models, whereas image segmentation is a popular sub-field [11,17,18]. The aim of medical image segmentation (MIS) is the automated identification and labeling of regions of interest (ROI) e.g. organs like lungs or medical abnormalities like cancer and lesions. In recent studies, medical image segmentation models based on neural networks proved powerful prediction capabilities and achieved similar results as radiologists regarding the performance [11,19]. It would be a helpful tool to implement such an automatic segmentation for COVID-19 infected regions as clinical decision support for physicians. By automatic highlighting abnormal features and ROIs, image segmentation is able to aid radiologists in diagnosis, disease course monitoring, reduction of time-consuming inspection processes and improvement of accuracy [11,12,20]. Nevertheless, training accurate and robust models requires sufficient annotated medical imaging data. Because manual annotation is labor-intensive, time-consuming and requires experienced radiologists, it is common that publicly available data is limited [11,12,16]. This lack of data often results in an overfitting of the traditional data-hungry models. Especially for COVID-19, large enough medical imaging datasets are currently unavailable [12,16].

In this work, we push towards creating an accurate and state-of-the-art MIS pipeline for COVID-19 lung infection segmentation, which is capable of being trained on small datasets consisting of 3D CT volumes. In order to avoid overfitting, we exploit extensive on-the-fly data augmentation, as well as diverse preprocessing methods. In order to further reduce the risk of overfitting, we implement the standard U-Net architecture instead of other more computational complex variants, like the residual architecture of the U-Net. Furthermore, we use a 5-fold cross-validation for reliable performance evaluation.

## 2. RELATED WORK

Since the breakthrough of convolutional neural network (CNN) architectures for computer vision, neural networks became one of the most accurate and popular machine learning algorithm for automated medical image analysis [11,17,21]. Two of the major tasks in this field are classification and segmentation. Whereas medical image classification aims to label a complete image to predefined classes (e.g. to a diagnosis), medical image segmentation aims to label each pixel in order to identify ROIs (e.g. organs or medical abnormalities). Popular deep learning architectures, which achieved performance equivalent to humans, are Inception-v3, ResNet, as well as DenseNet for classification and VB-Net, U-Net and various variants of the U-Net for segmentation [12,22–24].

In reaction to the rapid spread of the coronavirus, many scientists quickly reacted and developed various approaches based on deep learning to contribute to the efforts against COVID-19. Furthermore, the scientific community focused their efforts on the development of models for COVID-19 classification, because x-ray and CT images of infected patients could be collected without further required annotations [12,20]. These classification algorithms can be categorized through their objectives: 1) Classification of COVID-19 from non-COVID-19 (healthy) patients, which resulted into models achieving a sensitivity of 94.1%, specificity of 95.5%, and AUC of 0.979 (*Jin et al*) [25]. 2) Classification of COVID-19 from other pneumonia, which resulted in models achieving a sensitivity of 90%, specificity of 96%, and AUC of 0.96 (*Li et al*) [26]. 3) Severity assessment of COVID-19, which resulted in a model achieving a true positive rate of 93.3%, true negative rate of 74.5%, and accuracy of 87.5% (*Tang et al.*) [27].

In the last weeks, clinicians started to publish COVID-19 CT images with annotated ROIs, which allowed the training of segmentation models. Automated segmentation is highly desired as COVID-19 application [12,28]. The segmentation of lung, lung lobes and lung infection provides accurate quantification data for progression assessment in follow-up, comprehensive prediction of severity in the enrollment and visualization of lesion distribution using percentage of infection (POI) [12]. Still, the limited amount of annotated imaging data causes a challenging task for detecting the variety of shapes, textures and localizations of lesions or nodules. Nonetheless, multiple approaches try to solve these problems with different methods. The most popular network models for COVID-19 segmentation are variants of the U-Net which achieved reasonable performance on sufficiently sized 2D datasets [5,12,29–33]. In order to compensate limited dataset sizes, more attention has been drawn to semi-supervised learning pipelines [12,34]. These methods optimize a supervised training on labeled data along with an unsupervised training on unlabeled data. Another approach is the development of special neural network architectures for handling limited dataset sizes. Frequently, attention mechanism are built into the classic U-Net architecture like the Inf-Net from *Fan et al.* [34] or the MiniSeg from *Qiu et al* [35]. Particularly worth mentioning is the development of a benchmark model with a 3D U-Net from *Ma et al*, because the authors also provide high reproducibility through a publicly available dataset [16,36].

## 3. MATERIALS AND METHODS

The implemented medical image segmentation pipeline can be summarized in the following core steps and is illustrated in figure 2:
- Dataset: 20x COVID-19 CT volumes
- Limited dataset → Utilization as variation database
- Several preprocessing methods
- Extensive data augmentation
- Patchwise analysis of high-resolution images
- Utilization of the standard 3D U-Net
- Model fitting based on Tversky index & cross-entropy
- Model predictions on overlapping patches
- 5-fold cross-validation via Dice similarity coefficient

This pipeline was based on MIScnn [37], which is an in-house developed open-source framework to setup complete medical image segmentation pipelines with convolutional neural networks and deep learning models on top of Tensorflow/Keras [38]. MIScnn supports extensive preprocessing, data augmentation, state-of-the-art deep learning models and diverse evaluation techniques.

### 3.1 Dataset of COVID-19 Chest CTs

In this study, we used the public dataset from *Ma et al.* which consists of 20 annotated COVID-19 chest CT volumes [16,36]. At the time of this paper, this dataset is the only publicly available 3D volume set with annotated COVID-19 infection segmentation [16]. The CT scans were collected from the Coronacases Initiative and Radiopaedia and were licensed under CC BY-NC-SA. Each CT volume was first labeled by junior annotators, then refined by two radiologists with 5 years of experience and afterwards the annotations verified by senior radiologists with more than 10 years of experience [16]. Despite the fact that the sample size is rather small, the annotation process led to an excellent high-quality dataset. The volumes had a resolution of 512x512 (Coronacases Initiative) or 630x630 (Radiopaedia) with a number of slices of about 176 by mean (200 by median). The CT images were labeled into four classes: Background, lung left, lung right and COVID-19 infection.

In our pipeline, we performed a 5-fold cross-validation on the dataset. This resulted in five fitting and inference runs with each time 16 samples as training dataset and 4 samples for prediction. We decided not to follow the convention of splitting the dataset into training, validation and testing sets due to the limited dataset size and because we do not configure any hyper parameters afterwards on basis of validation/testing results.

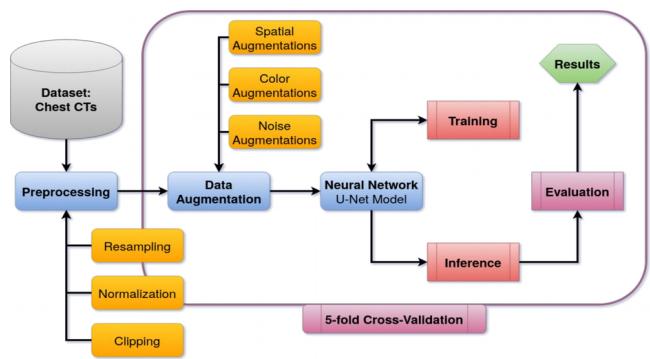

Figure 2: Flowchart diagram of the implemented medical image analysis pipeline for COVID-19 lung infection segmentation. The workflow is starting with the COVID-19 dataset and ending with the computed evaluation results for each fold in the cross-validation.

### 3.2 Preprocessing

In order to simplify the pattern finding and fitting process for the model, we applied several preprocessing methods on the dataset.

We exploited the Hounsfield units (HU) scale by clipping the pixel intensity values of the images to -1,250 as minimum and +250 as maximum, because we were interested in infected regions (+50 to +100 HU) and lung regions (-1,000 to -700 HU). It was only possible to apply the clipping approach on the Coronacases Initiative CTs, because the Radiopaedia volumes were already normalized to a grayscale range between 0 and 255.

Varying signal intensity ranges of images can drastically influence the fitting process and the resulting performance of segmentation models [39]. For achieving dynamic signal intensity range consistency, it is recommended to scale and standardize imaging data. Therefore, we normalized the Coronacases Initiative CT volumes likewise to grayscale range. Afterwards, all samples were standardized via z-score.

Medical imaging volumes have commonly inhomogeneous voxel spacings. The interpretation of diverse voxel spacings is a challenging task for deep neural networks. Therefore, it is possible to drastically reduce complexity by resampling volumes in an imaging dataset to homogeneous voxel spacing, which is also called target spacing. Resampling voxel spacings also directly resizes the volume shape and determines the contextual information, which the neural network model is able to capture. As a result, the target spacing has a huge impact on the final model performance. We decided to resample all CT volumes to a target spacing of 1.58x1.58x2.70, resulting in a median volume shape of 267x254x104.

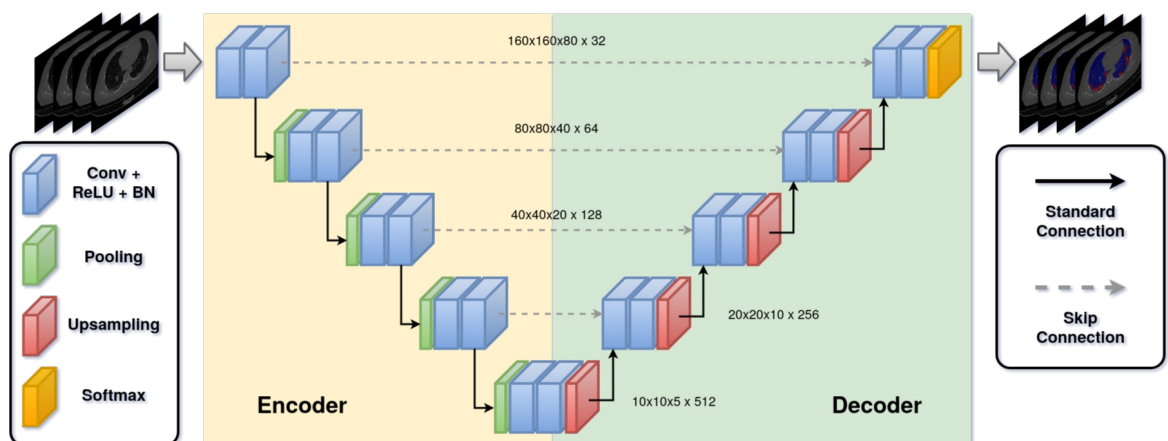

Figure 3: The architecture of the standard 3D U-Net. The network takes a 3D patch (cuboid) and outputs the segmentation of lungs and infected regions by COVID-19. Skip connections were implemented with concatenation layers. Conv: Convultional layer; ReLU: Rectified linear unit layer; BN: Batch normalization.

### 3.3 Data Augmentation

The aim of data augmentation is to create more data of reasonable variations of the desired pattern and, thus, artificially increase the number of training images. In order to compensate the small dataset size, we performed extensive data augmentation by using the batchgenerators interface within MIScnn. The batchgenerators package is an API for state-of-the-art data augmentation on medical images from the Division of Medical Image Computing at the German Cancer Research Center (DKFZ) [40]. We implemented three types of augmentations: Spatial augmentation by mirroring, elastic deformations, rotations and scaling. Color augmentations by brightness, contrast and gamma alterations. Noise augmentations by adding Gaussian noise. We performed the data augmentation on-the-fly on each image before it was forwarded into the neural network model. Furthermore, each augmentation method had a random probability of 15% to be applied on the current image with random intensity or parameters (e.g. random angle for rotation). Through this technique, the probability that the model encounters the exact same image twice during the training process decreases significantly.

### 3.4 Patchwise Analysis and Batch Generation

In image analysis there are two popular methods: The analysis of full images or patchwise by slicing the volume into smaller cuboid patches [11]. We selected the patchwise approach in order to exploit random cropping for the fitting process. Through random forwarding only a single cropped patch from the image to the fitting process, another type of data augmentation is induced, and the risk of overfitting additionally decreased. Furthermore, full image analysis requires unnecessary resolution reduction of the 3D volumes in order to handle the enormous GPU memory requirements. By slicing the volumes into patches with a shape of 160x160x80, we were able to utilize high-resolution data.

For inference, the volumes were sliced into patches according to a grid. Between the patches, we introduced an overlap of half the patch size (80x80x40) to increase prediction performance. After the inference of each patch, they were reassembled into the original volume shape, whereas overlapping regions were averaged.

The complete batch generation process, including the patch cropping and data augmentation for training, was implemented as on-the-fly. This means that batches are created during the fitting process instead beforehand. This allowed the creation of novel and unique images by the data augmentation in each iteration. For training, we used a batch size of 2.

### 3.5 Neural Network Model

The neural network architecture and its hyper parameters are one of the key parts in a medical image segmentation pipeline. In this work, we implemented the standard 3D U-Net as architecture in order to avoid unnecessary parameter increase by more complex architectures like the residual variant of the 3D U-Net [23,41,42]. Upsampling was achieved via transposed convolution and downsampling via maximum pooling. The architecture used 32 feature maps at its highest resolution and 512 at its lowest. All convolutions were applied with a kernel size of 3x3x3 in a stride of 1x1x1, except for up- and downsampling convolutions which were applied with a kernel size of 2x2x2 in a stride of 2x2x2.

In medical image segmentation, it is common that semantic annotation includes a strong bias in class distribution towards the background class. Our dataset revealed a class distribution of 89% for background, 9% for lungs and 1% for infection. In order to compensate this class bias, we utilized the sum of the Tversky index [43] and the categorical cross-entropy as loss function for model fitting (1).

$$L_{total} = L_{Tversky} + L_{CCE} \qquad (1)$$

$$L_{Tversky} = N - \sum_{c=1}^{N} \frac{TP_c}{TP_c + \alpha \cdot FN_c + \beta \cdot FP_c} \qquad (2)$$

$$L_{CCE} = -\sum_{c=1}^{N} y_{o,c} \log(p_{o,c}) \qquad (3)$$

We implemented a multi-class adaptation for the Tversky index (2), which is an asymmetric similarity index to measure the overlap of the segmented region with the ground truth. It allows for flexibility in balancing the false positive rate (*FP*) and false negative (*FN*) rate. The cross-entropy (3) is a commonly used loss function in machine learning and calculates the total entropy between the predicted and true distribution. The multi-class adaptation for multiple categories (categorical cross-entropy) is represented through the sum of the binary cross-entropy for each class c, whereas $y_{o,c}$ is the binary indicator whether the class label *c* is the correct classification for observation *o*. The variable $p_{o,c}$ is the predicted probability that observation *o* is of class *c*.

For model fitting, an Adam optimization was used with the initial weight decay of 1e-3 [44]. We utilized a dynamic learning rate which reduced the learning rate by a factor of 0.1 in case the training loss did not decrease for 15 epochs. The minimal learning rate was set to 1e-5. In order to further reduce the risk of overfitting, we exploited the early stopping technique for training, in which the training process stopped without a fitting loss decrease after 100 epochs. The neural network model was trained for a maximum of 1000 epochs. Instead of the common epoch definition as a single iteration over the dataset, we defined an epoch as the iteration over 150 training batches. This allowed for an improved fitting process for randomly generated batches in which the dataset acts as a variation database. According to our available GPU VRAM, we selected a batch size of 2.

### 3.6 Evaluation Metrics

During the fitting process, we computed the segmentation performance for each epoch on randomly cropped and data augmented patches from the validation dataset. This allowed for an evaluation of the overfitting on the training data.

After the training, we used three widely popular evaluation metrics in the community for medical image analysis to do the inference performance measurement in order to measure the segmentation overlap between prediction and ground truth. The Dice similarity coefficient, defined in (4), is the most widespread metric in computer vision. In contrast, the sensitivity (5) and specificity (6) are one of the most popular metrics in medical fields. All metrics are based on the confusion matrix, where *TP*, *FP*, *TN* and *FN* represent the true positive, false positive, true negative and false negative rate, respectively.

$$DSC = \frac{2 \cdot TP}{2 \cdot TP + FP + FN} \qquad (4)$$

$$Sensitivity = \frac{TP}{TP + FN} \qquad (5)$$

$$Specificity = \frac{TN}{TN + FP} \qquad (6)$$

We calculated the evaluation metrics for each fold in the cross-validation, and thus, for all samples in our dataset. The two lung classes (lung left and lung right) were averaged by mean into a single class (lungs) during the evaluation.

### 3.7 Code Reproducibility

In order to ensure full reproducibility and to create a base for further research, the complete code of this project, including extensive documentation, is available in the following Git repository:
https://github.com/frankkramer-lab/covid19.MIScnn

### 4. RESULTS & DISCUSSION

The sequential training of the complete cross-validation on a single GPU took around 130 hours. All folds did not require the entire 1000 epochs for training and instead were early stopped after an average of 312 epochs.

Through validation monitoring during the training, no overfitting was observed. The training and validation loss function revealed no significant distinction from each other, which can be seen in figure 4. During the fitting, the performance settled down at a loss of around 0.383 which is a generalized DSC (average of all class-wise DSCs) of around 0.919. Because of this robust training process without any signs of overfitting, we concluded that fitting on randomly generated patches via extensive data augmentation and random cropping from a variant database, is highly efficient for limited imaging data.

After the training, the inference revealed a strong segmentation performance for lungs and COVID-19 infected regions, which is illustrated in figure 6. Overall, the cross-validation models achieved a DSC of around 0.956 for lung and 0.761 for COVID-19 infection segmentation. Furthermore, the models achieved a sensitivity and specificity of 0.956 and 0.998 for lungs, as well as 0.730 and 0.999 for infection, respectively. More details on the inference performance is listed in table 1. From a medical perspective, detection of COVID-19 infection is a challenging task and one of the reasons for the weaker segmentation accuracy in contrast to the lung segmentation. The reason for this is the variety of GGO and pulmonary consolidation morphology. Nevertheless, our medical image segmentation pipeline allowed fitting a model which is able to segment COVID-19 infection with state-of-the-art accuracy that is comparable to models trained on large datasets.

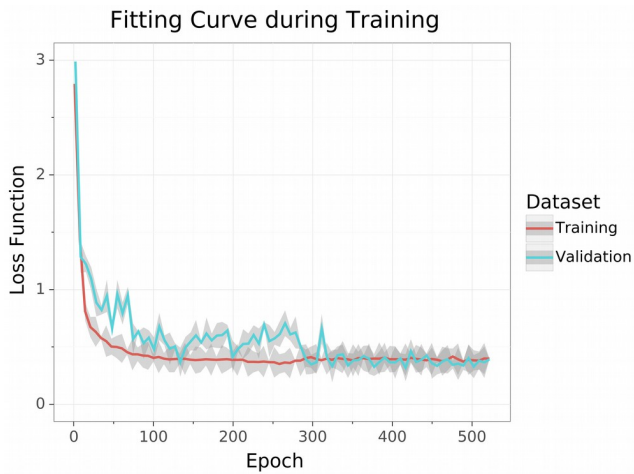

Figure 4: Loss course during the training process for training (red) and validation (cyan) data. The lines were computed via Gaussian Process Regression and represent the average loss across all folds. The gray areas around the lines represents the confidence intervals.

For further evaluation, we compared our pipeline to other available COVID-19 segmentation approaches based on CT scans. The authors (*Ma et al.*), who also provided the dataset we used for our analysis, implemented a 3D U-Net approach as a baseline for benchmarking [16]. They were able to achieve a DSC of 0.70355 and 0.6078 for lungs and COVID-19 infection, respectively. With our model, we were able to outperform this baseline. It is important to mention that we trained with a cross-validation distribution of 80% training and 20% testing, whereas they used the inverted distribution (20% training and 80% testing). Another approach from *Yan et al.* developed a novel neural network architecture (COVID-SegNet) specifically designed for COVID-19 infection segmentation with limited data [29]. The authors tested their architecture on a limited dataset consisting of ten COVID-19 cases from Brainlab Co. Ltd (Germany) and were able to achieve a DSC of 0.987 and 0.726 for lungs and infection, respectively. Hence, COVID-SegNet as well as our approach achieved similar results. This raises the question, if it is possible to further increase our performance by switching from the standard U-Net of our pipeline to an architecture specifically designed for COVID-19 infection segmentation like COVID-SegNet. Further approaches, with the aim to utilize specifically designed architectures, were Inf-Net (*Fan et al.*) and MiniSeg (*Qiu et al.*) [34,35]. Both were trained on 2D CT scans and achieved for COVID-19 infection segmentation DSCs of 0.764 and 0.773, respectively. Although diverse datasets were used for training, which leads to incomparability of the results, it is highly impressive that they achieved similar performance as approaches based on 3D imaging data. The 3D transformation of these architectures and the integration into our pipeline would be an interesting experiment to evaluate improvement possibilities.

Table 1: Achieved results showing the median Dice similarity coefficient (DSC), the sensitivity (Sens) and specificity (Spec) on Lung and COVID-19 infection segmentation for each CV fold and the global average (AVG).

|      | **Lungs** |       |       | **COVID-19** |       |       |
|------|-------|-------|-------|-------|-------|-------|
| **Fold** | DSC   | Sens. | Spec. | DSC   | Sens. | Spec. |
| 1    | 0.907 | 0.913 | 0.995 | 0.556 | 0.447 | 0.999 |
| 2    | 0.977 | 0.979 | 0.998 | 0.801 | 0.875 | 0.999 |
| 3    | 0.952 | 0.945 | 0.999 | 0.829 | 0.796 | 0.999 |
| 4    | 0.979 | 0.975 | 0.999 | 0.853 | 0.836 | 0.999 |
| 5    | 0.967 | 0.967 | 0.999 | 0.765 | 0.697 | 0.999 |
| AVG  | 0.956 | 0.956 | 0.998 | 0.761 | 0.730 | 0.999 |

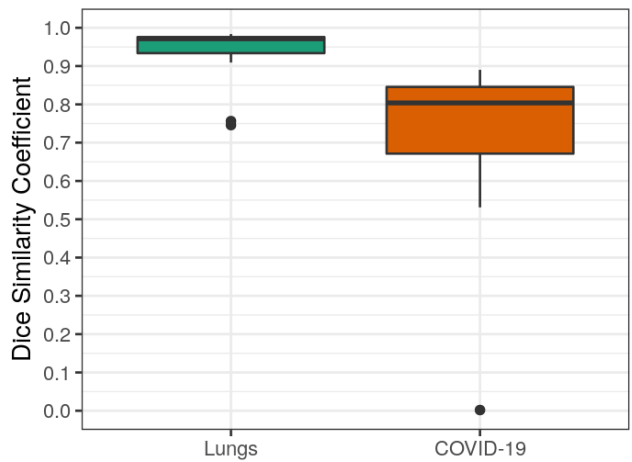

Figure 5: Box plots showing the result distribution from the 5-fold cross-validation on Lung and COVID-19 infection segmentation.

However, it is important to note that the majority of current segmentation approaches in research are not suited for clinical usage. The bias of current models is that they are only trained with COVID-19 related images. Therefore, it is not certain how good the models can differentiate between COVID-19 lesions and other pneumonia, or entirely unrelated medical conditions like cancer. Furthermore, identical to COVID-19 classification, the models reveal huge differences depending on which dataset they were trained on. Segmentation models purely based on COVID-19 scans are often not able to segment accurately in the presence of other medical conditions [16]. Additionally, there is a high potential for false positive segmentation of pneumonia lesions that are not caused by COVID-19. This demonstrates that these models could be biased and are not suitable for COVID-19 screening. Nevertheless, current infection segmentation models are already highly accurate for confirmed COVID-19 imaging. This offers the opportunity for quantitative assessment and disease monitoring as applications in clinical studies.

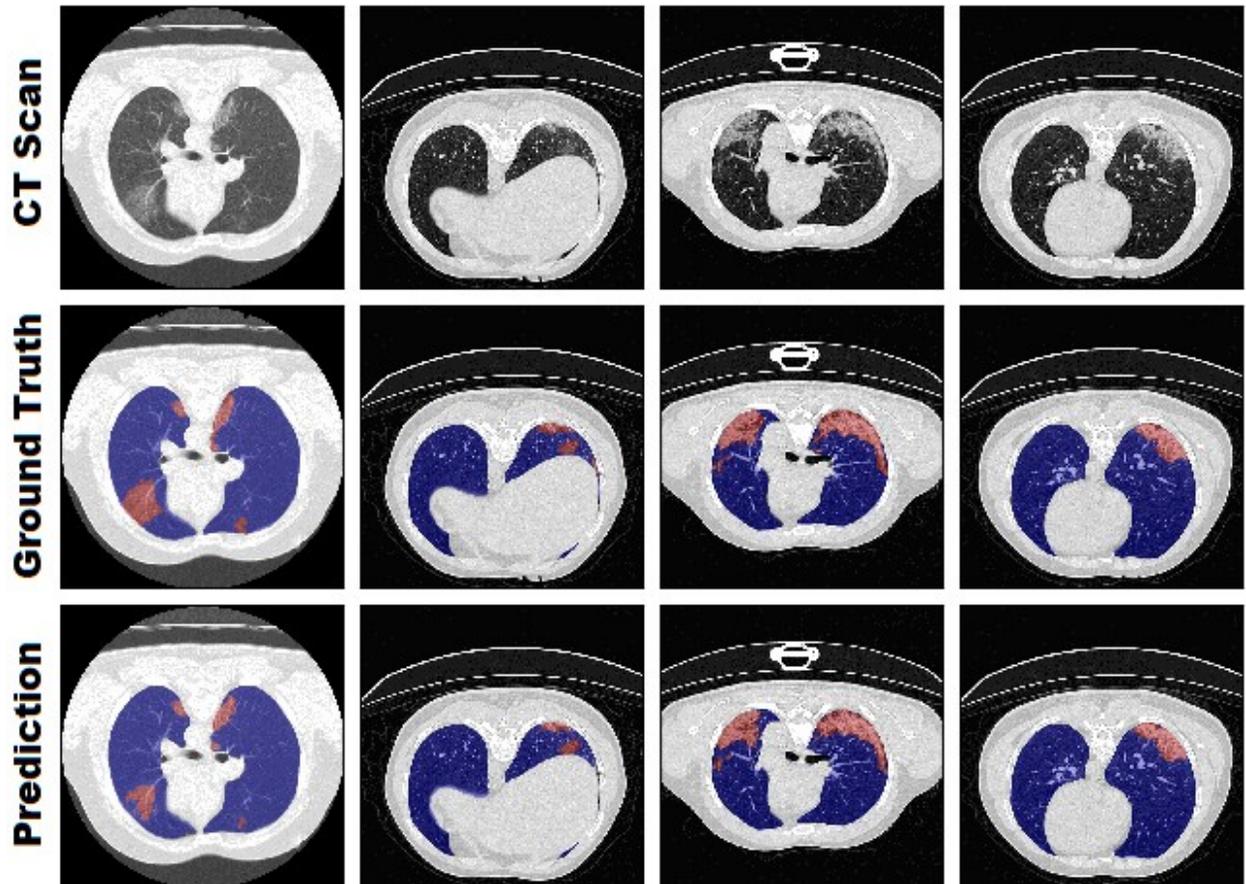

Figure 6: Visual comparison of the segmentation between ground truth and our model on four slices from different CT scans.

Despite that our model and those of others, which are based on limited data, are capable for accurate segmentation, it is essential to discuss their robustness. Currently, there are no large as well as annotated imaging datasets available for COVID-19 segmentation [16]. Existing small datasets may have incomplete and inaccurate labels which results in challenging handicaps for models. More imaging data with more variance (different COVID-19 states, other pneumonia, etc.) need to be collected, annotated and published for researchers. Similar to *Ma et al.*, community accepted benchmark datasets have to be established in order to fully ensure robustness as well as comparability of models [16,36].

## 5. CONCLUSION

In this paper, we developed and evaluated an approach for automated segmentation of COVID-19 infected regions in CT volumes. Our method focuses on on-the-fly generation of unique and random image patches for training by performing several preprocessing methods and exploiting extensive data augmentation. Thus, it is possible to handle limited dataset sizes which act as variant database. Instead of novel and complex neural network architectures, we utilized the standard 3D U-Net. We proved that our medical image segmentation pipeline is able to successfully train accurate and robust models without overfitting on limited data. Furthermore, we were able to outperform current state-of-the-art semantic segmentation approaches for lungs and COVID-19 infected regions. Our work has great potential to be applied as a clinical decision support system for COVID-19 quantitative assessment and disease monitoring in a clinical environment. Nevertheless, further research is needed on COVID-19 semantic segmentation in clinical studies for evaluating clinical performance and robustness.

## ACKNOWLEDGMENTS

We want to thank Bernhard Bauer and Fabian Rabe for sharing their GPU hardware (Nvidia Quadro P6000) with us which was used for this work. We also want to thank Dennis Klonnek, Jana Glöckler, Johann Frei, Florian Auer, Peter Parys, Zaynab Hammoud and Edmund Müller for their useful comments.

## FUNDING

This work is a part of the DIFUTURE project funded by the German Ministry of Education and Research (Bundesministerium für Bildung und Forschung, BMBF) grant FKZ01ZZ1804E.

## CONFLICTS OF INTEREST

The authors declare no conflicts of interest.

## REFERENCES

[1] Dong E, Du H, Gardner L. An interactive web-based dashboard to track COVID-19 in real time. Lancet Infect Dis 2020;20:533–4. doi:10.1016/S1473-3099(20)30120-1.

[2] W. H. O. Coronavirus disease (COVID-19) pandemic. 2020.

[3] Sohrabi C, Alsafi Z, O'Neill N, Khan M, Kerwan A, Al-Jabir A, et al. World Health Organization declares global emergency: A review of the 2019 novel coronavirus (COVID-19). Int J Surg 2020;76:71–6. doi:10.1016/j.ijsu.2020.02.034.

[4] RKI - Coronavirus SARS-CoV-2 - SARS-CoV-2 Steckbrief zur Coronavirus-Krankheit-2019 (COVID-19) n.d. https://www.rki.de/DE/Content/InfAZ/N/Neuartiges_Coronavirus/Steckbrief.html (accessed May 24, 2020).

[5] Amyar A, Modzelewski R, Ruan S. MULTI-TASK DEEP LEARNING BASED CT IMAGING ANALYSIS FOR COVID-19: CLASSIFICATION AND SEGMENTATION. MedRxiv Prepr 2020. doi:https://doi.org/10.1101/2020.04.16.20064709.

[6] Rodriguez-Morales AJ, Cardona-Ospina JA, Gutiérrez-Ocampo E, Villamizar-Peña R, Holguin-Rivera Y, Escalera-Antezana JP, et al. Clinical, laboratory and imaging features of COVID-19: A systematic review and meta-analysis. Travel Med Infect Dis 2020;34:101623. doi:10.1016/j.tmaid.2020.101623.

[7] Singhal T. A Review of Coronavirus Disease-2019 (COVID-19). Indian J Pediatr 2020;87:281–6. doi:10.1007/s12098-020-03263-6.

[8] Salehi S, Abedi A, Balakrishnan S, Gholamrezanezhad A. Coronavirus Disease 2019 (COVID-19): A Systematic Review of Imaging Findings in 919 Patients. Cardiopulm Imaging • Or Iginal Res 2020. doi:10.2214/AJR.20.23034.

[9] Ai T, Yang Z, Hou H, Zhan C, Chen C, Lv W, et al. Correlation of Chest CT and RT-PCR Testing in Coronavirus Disease 2019 (COVID-19) in China: A Report of 1014 Cases. Radiology 2020:200642. doi:10.1148/radiol.2020200642.

[10] Fang Y, Zhang H, Xie J, Lin M, Ying L, Pang P, et al. Sensitivity of Chest CT for COVID-19: Comparison to RT-PCR. Radiology 2020:200432. doi:10.1148/radiol.2020200432.

[11] Litjens G, Kooi T, Bejnordi BE, Setio AAA, Ciompi F, Ghafoorian M, et al. A survey on deep learning in medical image analysis. Med Image Anal 2017;42:60–88. doi:10.1016/j.media.2017.07.005.

[12] Shi F, Wang J, Shi J, Wu Z, Wang Q, Tang Z, et al. Review of Artificial Intelligence Techniques in Imaging Data Acquisition, Segmentation and Diagnosis for COVID-19. IEEE Rev Biomed Eng 2020:1–1. doi:10.1109/rbme.2020.2987975.

[13] Rubin GD, Ryerson CJ, Haramati LB, Sverzellati N, Kanne JP, Raoof S, et al. The Role of Chest Imaging in Patient Management during the COVID-19 Pandemic: A Multinational Consensus Statement from the Fleischner Society. Chest 2020. doi:10.1016/j.chest.2020.04.003.

[14] Strunk JL, Temesgen H, Andersen H, Packalen P. Imaging Profile of the COVID-19 Infection: Radiologic Findings and Literature Review Authors: 2014;80:1–8. doi:10.14358/PERS.80.2.000.

[15] Manning D, Ethell S, Donovan T, Crawford T. How do radiologists do it? The influence of experience and training on searching for chest nodules. Radiography 2006;12:134–42. doi:10.1016/j.radi.2005.02.003.

[16] Ma J, Wang Y, An X, Ge C, Yu Z, Chen J, et al. Towards Efficient COVID-19 CT Annotation: A Benchmark for Lung and Infection Segmentation 2020:1–7.

[17] Wang G. A perspective on deep imaging. IEEE Access 2016;4:8914–24. doi:10.1109/ACCESS.2016.2624938.

[18] Aggarwal P, Vig R, Bhadoria S, C.G.Dethe A. Role of Segmentation in Medical Imaging: A Comparative Study. Int J Comput Appl 2011;29:54–61. doi:10.5120/3525-4803.

[19] Lee K, Zung J, Li P, Jain V, Seung HS. Superhuman Accuracy on the SNEMI3D Connectomics Challenge 2017:1–11.

[20] Bullock J, Luccioni A, Pham KH, Lam CSN, Luengo-Oroz M. Mapping the Landscape of Artificial Intelligence Applications against COVID-19 2020:1–32.

[21] Shen D, Wu G, Suk H-I. Deep Learning in Medical Image Analysis. Annu Rev Biomed Eng 2017;19:221–48. doi:10.1146/annurev-bioeng-071516-044442.

[22] Jin S, Wang B, Xu H, Luo C, Wei L, Zhao W, et al. AI-assisted CT imaging analysis for COVID-19 screening: Building and deploying a medical AI system in four weeks. MedRxiv 2020:2020.03.19.20039354. doi:10.1101/2020.03.19.20039354.

[23] Ronneberger O, Philipp Fischer, Brox T. U-Net: Convolutional Networks for Biomedical Image Segmentation. Lect Notes Comput Sci (Including Subser Lect Notes Artif Intell Lect Notes Bioinformatics) 2015;9351:234–41. doi:10.1007/978-3-319-24574-4.

[24] He K, Zhang X, Ren S, Sun J. Deep Residual Learning for Image Recognition 2015.

[25] Jin C, Chen W, Cao Y, Xu Z, Zhang X, Deng L, et al. Development and Evaluation of an AI System for COVID-19 Diagnosis. MedRxiv 2020:2020.03.20.20039834. doi:10.1101/2020.03.20.20039834.

[26] Hou H, Lv W, Tao Q, Hospital T, Company JT, Ai T, et al. Artificial Intelligence Distinguishes COVID-19 from Community Acquired Pneumonia on Chest CT 2020;2020.

[27] Shi F, Xia L, Shan F, Wu D, Wei Y, Yuan H, et al. Large-Scale Screening of COVID-19 from Community Acquired Pneumonia using Infection Size-Aware Classification 2020.

[28] Gozes O, Frid-Adar M, Greenspan H, Browning PD, Bernheim A, Siegel E. Rapid AI Development Cycle for the Coronavirus (COVID-19) Pandemic: Initial Results for Automated Detection & Patient Monitoring using Deep Learning CT Image Analysis 2020.

[29] Yan Q, Wang B, Gong D, Luo C, Zhao W, Shen J, et al. COVID-19 Chest CT Image Segmentation -- A Deep Convolutional Neural Network Solution 2020:1–10.

[30] Chen X, Yao L, Zhang Y. Residual Attention U-Net for Automated Multi-Class Segmentation of COVID-19 Chest CT Images 2020;14:1–7.

[31] Gozes O, Frid-Adar M, Sagie N, Zhang H, Ji W, Greenspan H. Coronavirus Detection and Analysis on Chest CT with Deep Learning 2020:1–8.

[32] Gaál G, Maga B, Lukács A. Attention U-Net Based Adversarial Architectures for Chest X-ray Lung Segmentation 2020:1–7.

[33] Zhou T, Canu S, Ruan S. An automatic COVID-19 CT segmentation based on U-Net with attention mechanism 2020:1–14.

[34] Fan D-P, Zhou T, Ji G-P, Zhou Y, Chen G, Fu H, et al. Inf-Net: Automatic COVID-19 Lung Infection Segmentation from CT Scans 2020;2019:1–11.

[35] Qiu Y, Liu Y, Xu J. MiniSeg: An Extremely Minimum Network for Efficient COVID-19 Segmentation 2020:1–10.

[36] Jun M, Cheng G, Yixin W, Xingle A, Jiantao G, Ziqi Y, et al. COVID-19 CT Lung and Infection Segmentation Dataset 2020. doi:10.5281/zenodo.3757476.

[37] Müller D, Kramer F. MIScnn: A Framework for Medical Image Segmentation with Convolutional Neural Networks and Deep Learning 2019.

[38] Martín Abadi, Ashish Agarwal, Paul Barham, Eugene Brevdo, Zhifeng Chen, Craig Citro, et al. TensorFlow: Large-Scale Machine Learning on Heterogeneous Systems 2015.

[39] Roy S, Carass A, Prince JL. Patch based intensity normalization of brain MR images. Proc. - Int. Symp. Biomed. Imaging, 2013. doi:10.1109/ISBI.2013.6556482.

[40] Isensee F, Jäger P, Wasserthal J, Zimmerer D, Petersen J, Kohl S, et al. batchgenerators - a python framework for data augmentation 2020. doi:doi:10.5281/zenodo.3632567.

[41] Çiçek Ö, Abdulkadir A, Lienkamp SS, Brox T, Ronneberger O. 3D U-net: Learning dense volumetric segmentation from sparse annotation. Lect Notes Comput Sci (Including Subser Lect Notes Artif Intell Lect Notes Bioinformatics) 2016;9901 LNCS:424–32. doi:10.1007/978-3-319-46723-8_49.

[42] Zhang Z, Liu Q, Wang Y. Road Extraction by Deep Residual U-Net. IEEE Geosci Remote Sens Lett 2018. doi:10.1109/LGRS.2018.2802944.

[43] Seyed SSM, Erdogmus D, Gholipour A. Tversky loss function for image segmentation using 3D fully convolutional deep networks 2017.

[44] Kingma DP, Lei Ba J. Adam: A Method for Stochastic Optimization. 2014.